# Highly tunable moiré superlattice potentials in twisted hexagonal boron nitrides

Kwanghee Han[1], Minhyun Cho[1,2], Taehyung Kim[1], Seung Tae Kim[1], Suk Hyun Kim[1,3], Sang Hwa Park[4], Sang Mo Yang[4], Kenji Watanabe[5], Takashi Taniguchi[6], Vinod Menon[2], Young Duck Kim[1,3]*

[1]Department of Physics, Kyung Hee University, Seoul, 02447, Republic of Korea.

[2]Department of Physics, City College of New York, New York, NY, 10031, USA.

[3]Department of Information Display, Kyung Hee University, Seoul, 02447, Republic of Korea.

[4]Department of Physics, Sogang University, Seoul, 04107, Republic of Korea.

[5]Research Center for Electronic and Optical Materials, National Institute for Materials Science, 1-1 Namiki, Tsukuba, 305-0044, Japan.

[6]Research Center for Materials Nanoarchitectonics, National Institute for Materials Science, 1-1 Namiki, Tsukuba, 305-0044, Japan.

*Corresponding author E-mail: ydk@khu.ac.kr

Keywords: *twisted hexagonal boron nitride, moiré superlattice, twistronics, ferroelectricity, remote Coulomb potential*

**Moiré superlattice of twisted hexagonal boron nitride (hBN) has emerged as an advanced atomically thin van der Waals interfacial ferroelectricity platform. Nanoscale periodic ferroelectric moiré domains with out-of-plane potentials in twisted hBN allow the hosting of remote Coulomb superlattice potentials to adjacent two-dimensional materials for tailoring strongly correlated properties. Therefore, the new strategies for engineering moiré length, angle, and potential strength are essential for developing programmable quantum materials and advanced twistronics applications devices. Here, we demonstrate the realization of twisted hBN-based moiré superlattice platforms and visualize the moiré domains and ferroelectric properties using Kelvin probe force microscopy. Also, we report the KPFM result of regular moiré superlattice in the large area. It offers the possibility to reproduce uniform moiré structures with precise control piezo stage stacking and heat annealing. We demonstrate the high tunability of twisted hBN moiré platforms and achieve cumulative multi-ferroelectric polarization and multi-level domains with multiple angle mismatched interfaces. Additionally, we observe the quasi-1D anisotropic moiré domains and show the highest resolution analysis of the local built-in strain between adjacent hBN layers compared to the conventional methods. Furthermore, we demonstrate *in-situ* manipulation of moiré superlattice potential strength using femtosecond pulse laser irradiation, which results in the optical phonon-induced atomic displacement at the hBN moiré interfaces. Our results pave the way to develop precisely programmable moiré superlattice platforms and investigate strongly correlated physics in van der Waals heterostructures.**

**Introduction**

Moiré superlattice in twisted van der Waals (vdW) materials can be used as a robust quantum simulation platform for strongly correlated physics due to its high tunability of quantum many-body interactions by controlling the carrier density[1,2], strain[3], and twisted angle[4,5]. Since the moiré superlattice structure can introduce a periodic electrostatic Coulomb potential for engineering nearby vdW layers[6,7], the opportunity to make additional tunability became one of the most important challenges for twistronics. Hexagonal boron nitride (hBN) is a two-dimensional (2D) ultra-wide bandgap material (> 6 eV)[8] that can be easily obtained by exfoliation like graphene[9,10] and transition metal dichalcogenides. Therefore, hBN has been widely used as a passivation material to improve the quality of vdW heterostructures for high-performance devices and studies of intrinsic quantum properties[1,11,12]. For instance, it is used for encapsulating layers[1,13], improving charge carrier mobility[11], and tunneling barriers[14].

In addition, hBN itself has intriguing properties when two layers are slightly misaligned to form moiré superlattice structures. When boron (B) and nitrogen (N) atoms are aligned, the BN and NB dipoles are generated along the out-of-plane axis. Unlike AA' stacking order, AB/BA stacking has only one orientation of dipoles[15–17]. Therefore, the symmetry breaking along the out-of-plane direction creates a ferroelectric property with out-of-plane polarization[15]. It leads to periodic potential differences by local stacking orders[2]. Because of the high tunability of periodic potential, twisted hBN is a promising platform to control 2D materials when they are placed in the proximity[6,7].

Previous research showed the moiré superlattice systems with twisted TMDC materials can exhibit interesting phenomena with the interlayer exciton[18,19]. However, these systems are only possible with interlayer exciton because moiré structure is formed by each TMDC layer.

Compared to the other papers, our research shows the possibilities to control all kinds of materials with proximity in twisted hBN platform[20,21].

Here, we report strategies for engineering twisted hBN moiré superlattice platforms by incorporating angle mismatch, shear strain, and multiple ferroelectric interfaces. We present the high tunability of twisted hBN moiré platforms and the realization of the cumulative multi-ferroelectric polarization domains with multiple angle mismatch interfaces. We also realized quasi-1D moiré superlattice potentials with applied strains. By analyzing the 1D stripe moiré pattern, we deduce the precise strain between two layers. In addition, we demonstrate a technique for *in-situ* tailoring of moiré superlattice with optical phonon-induced lattice deformations generated by femtosecond laser. By using these methods, the twisted hBN moiré superlattice platform can offer opportunities to realize programmable moiré quantum materials.

## 2. Results
### 2.1. Twisted hBN moiré superlattice

Twisted hBN possesses six different high-symmetry stacking configurations with two groups called "parallel" and "antiparallel"[22]. In the antiparallel case, the AA' stacking order can be realized with columns of alternating boron and nitrogen atoms repeating. Bulk hBN usually has the AA' stacking order because it is the most stable registry according to their formation energy. For the parallel twist orientation, by contrast, stacking configurations can be AA, AB, or BA (Fig. 1a)[16]. In AB and BA stacking orders, the interaction by pairs of boron and nitrogen atom (BN or NB) results in out-of-plane dipoles, and it creates a ferroelectric property breaking the inversion symmetry[17]. Due to the charge redistribution, the electrostatic potential will have different values in the AB versus BA stacking order. Unlike these orders, the AA stacking is unstable because two nitrogen (boron) atoms are stacked atop each other[22]. To minimize the interlayer configurations energy, the atomic structures are reconstructed as triangular repetitive moiré patterns, as shown in Fig. 1b[16,23].

In this work, twisted hBN samples were fabricated by using the 'tear & stack method' [1,16,24–26] to make the hBN layer on the other layer with a precise twist angle alignment between two layers. In addition, we performed vacuum thermal annealing to remove the polymer residue and enhance the strain relaxation between two hBN layers[27,28]. To confirm the ferroelectric moiré domains in the twisted hBN, we measured the surface potential differences between AB and BA stacking of hBN using Kelvin probe force microscopy (KPFM) as shown in Fig. 1b[15,23].

Figure 1c shows an atomic force microscope (AFM) topography image of a representative twisted hBN sample with an atomically flat surface without any bubbles, ripples, or polymer residues. The surface potential ($V_S$) is obtained by doing an AFM of the same area in KPFM mode. Compared to the previous research[15,23], our result shows the regular moiré patterns in the largest area because of the precise piezo control of transfer stage, and heat annealing. The KPFM image in Fig. 1d. shows a large area of regular moiré domains from the sample with moiré area bigger than ~123 μm². This result shows the possibility of constructing large-area high-quality moiré superlattice structures using the tear & stack method. For twisted triangular lattices, the moiré length is $\lambda_m(\theta) = a/2\sin\left(\frac{\theta}{2}\right)$, where $\theta$ is the twist angle between hBN layers and $a = 2.504$ Å is the lattice constant of the hBN[29,30]. Therefore, the size of the moiré length dramatically increases when the twist angle gets smaller. We note that since the sample has a strain induced by topographic step difference, the twist angle gradually changes as seen in the bottom right corner of Fig. 1d.

Figure 1e shows KPFM images cropped from Fig. 1d. To analyze the relationship between the $\lambda_m$ and the surface potential difference ($\Delta V_S$), we choose two different areas encircled by the red and blue rectangle. A cropped KPFM image in the red (blue) encircled area exhibits the smaller (bigger) moiré pattern. In Fig. 1f, there are two graphs of the potential line profile extracted from the red (blue) line. It demonstrates the $\lambda_m \sim 90$ nm ($\sim 260$ nm), and the potential depth is $\Delta V_S \sim 157$ mV ($\sim 269$ mV). According to the calculation, the twist angle of the sample is changed from $\sim 0.16°$ to $\sim 0.06°$. Thus, the potential depth is larger when the twist angle is smaller, and the potential depth saturates while the moiré length increases.

Figure 1g is the statistical analysis of the relation between the moiré length and the potential depth. The yellow dots are from the experimental data in Fig. 1d, and a blue (sky-blue) line is a fit. By calculating the electrical polarization $P$, we can examine the electrostatic potential at the twisted hBN surface using[6]

$$V(\mathbf{R}, z) \approx sgn(z) \frac{P(\mathbf{R})}{2\epsilon_0} e^{-G|z|}, \quad (1)$$

where $P(\mathbf{R}) = \int z' \Delta \rho(\mathbf{R}, z') dz'$ and $G = 4\pi/(\sqrt{3}\lambda_m)$. $\mathbf{R}$ is the lateral position vector, $z$ is the vertical distance to the interface, and $\epsilon_0$ is the vacuum permittivity. The $z$ is the sum of top hBN thickness, $d_T = 1.7$ nm, and tip-to-surface distance, $z_{lift} = 12.1\ nm$. (see Table S1 in Supporting Information) So, $z = 13.8\ nm$ was used for fitting curve and the result is well-matched with experiment data. Therefore, the relation between moiré length and potential depth can be fitted by $\Delta V \sim exp(-\frac{4\pi z}{\sqrt{3}\lambda_m})$, resulting in the potential depth increasing when $\lambda_m$ gets longer and saturating at some point. All the details of the fitting curve are included in the Supporting Note 8.

## 2.2. Cumulative hBN moiré superlattice potential

Recent research shows that cumulative moiré interfaces can make more polarization states[24,31]. So, we observed the surface potential with KPFM to explore the twisted hBN with additional moiré interfaces. Figure 2a presents a schematic of twisted hBN with two different superlattice interfaces. By precisely aligning an additional hBN layer atop twisted hBN, a new moiré interface can be overlapped onto the original moiré interface. If two interfaces have different twist angles from each other, the multi-level surface potential can be easily observed due to the combination of polarizations from the two interfaces. Figure 2b is an AFM topography image of a representative sample. In Figure 2c, the corresponding KPFM image shows the moiré pattern by two superlattice interfaces. Along the white dashed lines, the big moiré pattern exhibits the surface potential difference. In addition, the smaller moiré patterns by another moiré interface are uniformly constructed. Since we know the moiré length from both interfaces, their respective twist angles can be calculated. The first superlattice interface has a moiré length with $\lambda_1 \simeq 350$ nm, and its twist angle is $\theta_1 \simeq 0.04°$. The second interface has $\lambda_2 \simeq 80$ nm, and $\theta_2 \simeq 0.2°$. Extracting a line profile along the red line in Fig. 2c, there is a step-like decrease of ~133.5 mV superimposed on the potential modulation of ~70 mV (Fig. 2d). The black arrows in the graph describe the potential modulation directions by polarization states from the two interfaces. The potential with two opposite polarizations (↑↓, ↓↑) has intermediate potential values. By contrast, the superimposed case with parallel polarizations (↑↑, ↓↓) shows an increased surface potential difference.

To compare a single moiré superlattice interface and the multiple interface structures, we observed the twisted hBN partially covered by an additional layer. Figure 2e is a schematic of the sample, which exhibits both single interface and double interfaces simultaneously. In Fig. 2f, white dashed lines describe the additional ~1.5 nm thick hBN layer on the twisted sample. (see

Table S1 in Supporting Information) The KPFM image in Fig. 2g shows the big moiré patterns constructed by the additional hBN layer encircled by white dashed lines. The small moiré patterns from the bottom interface are uniformly distributed on the sample. Figure 2h is a potential line profile from the red line. In the right-side of the graph, the amplitude of potential modulation is ~123 mV without an additional layer. The black arrows represent a single orientation of polarization (↑, ↓). But on the additional layer area, the multi-polarization states (↑↑, ↑↓, ↓↓, ↓↑) can be realized due to the potential modulation from another superlattice interface. So, the average potential increases to +16.3 mV and decreases to -89.5 mV by potential modulation to the upwards and downwards.

So far, the twisted vdW platform has a limitation in scaling the polarization, and only two states were switchable[2,15,16,23]. However, the multi-level polarization states can be realized by introducing an additional moiré interface[7,24]. Also, the polarizations can be tuned by the hBN thickness[31]. In addition, the twisted hBN platform offers a more stable system since it has been used for encapsulating other materials to realize the high-performance optoelectronic device which shows the intrinsic quantum properties[1,11–13,32–34]. Therefore, introducing an additional moiré interface with hBN could be key to realizing a more stable and cumulative multi-ferroelectric scalable platform.

**2.3. Strained induced anisotropic moiré domains**

There are four types of moiré lattices such as twist, isotropic strain, diagonal pure shear, and horizontal simple shear (HSS)[35]. The twisted 2D homobilayer is the most well-known type of moiré superlattice which forms triangular patterns when it is reconstructed. Isotropic strain and pure shear case also have triangular moiré patterns after atomic relaxation. By contrast, the HSS where the layers are subject to horizontal uniaxial simple shear strain has one-dimensional moiré in the vertical direction[35].

We obtained KPFM and PFM (Piezoelectric Force Microscopy) results of 1D moiré superlattice by inducing HSS on the twisted hBN sample. A simple schematic of HSS in Fig. 3a shows two identical hexagonal layers are sheared by strain to the horizontal direction. The red dots and blue dots represent the atoms in two adjacent layers respectively. As one layer sheared to left side and the other layer to right side, the slight misalignment will make the local stacking order change. The configuration in Fig. 3b shows the quasi-one-dimensional (1D) moiré superlattice after atomic reconstruction. The moiré pattern is elongated by strain and forms anisotropic moiré potential. As shown in Fig. 3b, the moiré superlattice pattern goes with AA/BA/AB stacking orders repetitively. The moiré length of HSS is calculated by $\lambda_M = a/\varepsilon$, where $a$ is the hBN lattice constant and $\varepsilon$ is the strain parameter[35].

To study the quasi 1D HSS moiré superlattice, we prepared the HSS hBN sample with two superlattice interfaces sheared vertically and horizontally, respectively. In the 3-dimensional schematic of the sample, the first superlattice interface constructed by a shear strain $\varepsilon_1$ have a striped shape moiré pattern in the vertical direction. By introducing an additional superlattice interface, the strain $\varepsilon_2$ can realize another moiré pattern in the horizontal direction as shown in the Fig. 3c. Figure 3d shows AFM topography of HSS hBN sample fabricated by folding and placing another layer with alignment. Due to the topographic difference induced by this method, simple shears along two different axes make moiré pattern elongated in each direction. In the KPFM image of Fig. 3f, there are horizontal and vertical superlattices at the same time.

To improve the resolution of visualization of the HSS moiré superlattice domain, we measured PFM with the Dual Frequency Resonance Tracking (DFRT) technique (see Methods)[36]. Since the PFM amplitude can visualize the domain walls[37], we obtained moiré domain walls in the PFM image (Fig. 3h). Due to this method, we could enhance the signal-to-noise ratio. By extracting line profile, we observed a clear transition of PFM amplitude (Fig. 3i). The DFRT-PFM result confirms the HSS moiré patterns in vertical and horizontal direction by visualizing domain walls. Since the strain of the 2D materials can induce the distortion of the moiré pattern by its elongation, the anisotropic potential with 1D stripes can be obtained and allow the accurate estimate of the nanoscale local strains in moiré superlattice platforms. The representative potential line profiles were extracted with red and blue lines respectively. The line profiles described in Fig. 3g and 3i show the moiré length. According to the analysis, the first moiré superlattices have a moiré length $\lambda_{m1} \simeq 240$ nm and its calculated strain is $\varepsilon_1 \simeq 0.104$ %. Meanwhile, the second moiré superlattices have $\lambda_{m2} \simeq 320$ nm and its strain is $\varepsilon_2 \simeq 0.078$ %.

Previous studies have primarily relied on conventional optical measurements to analyze strain or twist angles. For instance, Raman spectroscopy and infrared (IR) optical microscopy have been employed to detect strain in hBN[38], while second harmonic generation (SHG) has been used to determine twist angles[4,39,40]. However, these conventional techniques have inherent limitations in accurately analyzing twist angles and strain in individual nanoscale moiré patterns. Optical measurements typically have optical beam sizes ranging from 1 to 10 μm, which can obscure nanoscale detection. Furthermore, SHG measurements are challenging in distinguishing clean intensity differences at small twist angles and resolving the continuous variation in angles and strains. In contrast, KPFM of moiré patterns offers a significant advantage by providing high-resolution analysis of twist angles (θ ~ 0.01°) and strain (ε ~ 0.01 %), which allows us to accurately resolve the nanoscale continuous variation of angle and strain in 2D van der Waals heterointerfaces. In addition, we note that controlling the various shear and strain will provide new design tools for the realization of moiré superlattice with different sizes and geometry[41].

**2.4. *in-situ* engineering of moiré superlattice potential strength**

hBN exhibits efficient deep ultraviolet (DUV) light emission due to the strong electron-phonon interaction induced radiative recombination[8,42], resulting in transverse optical (TO) or longitudinal optical (LO) phonons that can be induced by femtosecond laser irradiation[43]. When optical excitation energy exceeds the indirect band gap of multilayer hBN (5.95 eV), it leads to significant electron-phonon interactions induced TO and LO emissions in hBN[42,44]. So, the interlayer mechanical shear and breathing mode between adjacent layers occurs by phonon emissions[45,46]. In addition, a recent study proposed a method to cleave hBN flakes by driving intense TO ($E_{1u}$) phonon resonance with a mid-IR femtosecond pulse laser[47]. The ferroelectric switching by electric fields[23,48,49] is complicated because the electric field should be applied to out-of-plane direction while the sliding of stacking order is in-plane direction. Also, an additional conducting layer is necessary to apply an electric field to the system. However, the in-plane phonon modes induced by pulse laser are much easier sources to program the ferroelectric switching. Moreover, the polarization switching of twisted hBN by ultrafast optical or electrical pulsed induced interlayer sliding in van der Waals moiré system has been theoretically predicted recently[50,51]. Therefore, optical phonon-induced interlayer shear modes with deep UV femtosecond pulse laser can offer *in-situ* engineering of moiré superlattice with deformation of twisted hBN lattice structure.

To demonstrate the *in-situ* engineering of moiré superlattice potential strength in twisted hBN, we irradiated a deep UV (193 nm, 6.42 eV) femtosecond pulse laser to the twisted hBN moiré platform, as shown in Fig. 4a. Figure 4b presents the KPFM image of a twisted hBN sample and regular moiré superlattice potentials before femtosecond laser irradiation. By extracting the line profile from the blue line from Fig. 4b, the amplitude of modulation is ~ 31.7 mV and the moiré length is ~ 220 nm (Fig. 4d). After femtosecond pulse laser irradiation with irradiance of 3.14 mJ/cm$^2$, the moiré patterns become blurry as shown in Fig. 4c. By extracting the line profile from the red line in Fig. 4c, the potential amplitude decreased to ~ 10.8 mV as shown in Fig. 4d. Compared to the original result, the moiré potential amplitude of modulation by ultrashort laser irradiation is diminished to $\Delta V$ ~ 20.9 mV.

Figure 4e presents the representative statistical comparison of moiré potential strengths before and after DUV femtosecond laser irradiation as shown in Fig 4b and c. According to the normalized number of counts for ferroelectric polarization strength ($V_S$), the average amplitude of moiré potential was reduced from ~ 32.4 mV to ~ 16.3 mV. Here, we note that the several effects of ferroelectric domain depolarization in 2D materials by damage, doping, and etching are not involved and are confirmed by AFM topography and KPFM after laser irradiation (see Supporting Note 3). In addition, we irradiated the twisted hBN sample under high vacuum conditions (~10$^{-6}$ Torr) to minimize contamination and damage.

By calculating the electrical polarization as a function of in-plane displacement $r$ between two boron atoms in the nearest neighbor layer, we estimate the in-plane relocation of atoms in AB/BA stacking[6]. If the polarization decreases by ~ 1/3, the atoms will be displaced about $\pm$ 0.6 ~ 0.65 Å from the original position (see Supporting Note 2). Moreover, the atomic displacement of hBN by TO phonon resonance can be increased by using higher laser pump intensity[43]. We also observed that increased laser irradiance enhances the modulation of moiré superlattice potentials in twisted hBN moiré platforms (see Supporting Note 4). Our result implied that ultrafast femtosecond laser allows the *in-situ* precise tailoring of moiré superlattice domains and optical controls of symmetries in twisted quantum materials[52]. Utilizing a femtosecond laser to modulate the moiré superlattice potential strength offers a significant advancement in the engineering of moiré based twistronics devices. This approach provides a superior alternative to conventional mechanical manipulation, which often leads to unpredictable complications such as strain, wrinkles, and tears.

## 3. Conclusion

In conclusion, we have demonstrated the twisted hBN-based moiré superlattice platforms and their various interfacial ferroelectric properties. First, we confirm that the moiré potential depth is related to the moiré length, and it could be affected by strain. In addition, we obtain the regular moiré patterns results in the large area with small variation. Furthermore, cumulative interfacial polarization with superimposed moiré interfaces can offer more scalable and switchable platforms. By inducing strain to the mis-aligned hBN, we found a different type of moiré structure with 1D stripe pattern with KPFM and DFRT-PFM. Controlling the strain of twisted hBN moiré superlattice allows us to engineer the excitons and polaritons in the adjacent TMDC layers with 1D stripe repetitive potentials. In addition, the observation of 1D moiré patterns offers the highest resolution to sense the lattice distortion and strain effects of twisted vdW materials. Lastly, programming moiré structure and potential with femtosecond laser will pave the way to *in-situ* control of moiré superlattice properties and quantum materials. We expect that further research to improve the tunability of twisted hBN platforms will provide comprehension for advanced 2D quantum optoelectronic application devices and realize future twistronics applications.

**Methods**

*Sample preparation*: All the samples were fabricated by a dry-transfer technique[1,25]. 1-18 nm thick hexagonal boron nitrides are exfoliated on $SiO_2$/Si substrates and proper flakes were selected when it is thin and flat with no wrinkle or tape residue (see Supporting Note 1). Two hBN flakes with the same angle orientation were aligned and stacked to construct a moiré superlattice with a small twist angle. We used the hemispherical substrate by depositing polydimethylsiloxane (PDMS) droplet onto a transparent slide glass. The substrate was covered with an adhesive polymer like polypropylene carbonate (PPC), or poly (bisphenol A carbonate) (PC). After picking up the first half, we used the automatic transfer stage to control the precise position and twist angle and overlapped to another piece. To make two interface samples, this step was repeated one more time. Finally, the stack was released onto a high-quality $SiO_2$/Si chip. The temperatures for PPC (PC) film to pick up and transfer were ∼ **40** ℃ (∼ **90** ℃) and ∼ **100** ℃ (∼ **180** ℃) respectively. To remove the film, we placed samples in the chloroform overnight and performed heat annealing at **300**∼**500** ℃ for 2 hours under a vacuum of $10^{-6}$ Torr.

*KPFM measurements*: KPFM measurements were performed using Park Systems XE100, NX10, and Bruker Multimode 8 atomic force microscopy in non-contact scanning mode. The electrostatic signal was measured by a built-in lock-in-amplifier. We used NSC36/Cr-Au and PFQNE-AL tip. The resonance frequency of the tips was 65~300 kHz and the force constant was 0.6~2 N/m. The AC voltage was applied to the cantilever with an amplitude of 3-5 V and a frequency of 17 kHz. The surface potential can be obtained by controlling DC voltage with servo to nullifying the work function difference between the tip and sample. Images were acquired by the Park Systems XEI program and the data were analyzed using the Gwyddion program.

*DFRT PFM measurement*: PFM measurements were performed by a Park Systems NX10 in contact scanning mode. To improve a signal-to-noise ratio in the piezoresponses, we used one of the PFM techniques called Dual Frequency Resonance Tracking (DFRT). In PFM, a conductive tip scans the sample with applying AC voltage between the tip and the bottom gate of sample[53]. When the AC voltage is applied on the ferroelectric materials, the piezoelectricity instantly either expands or contracts the sample so that we can observe the polarization information. The standard PFM applies a low frequency AC voltage which is far from the contact resonance of the cantilever. However, the DFRT-PFM use two sidebands left and right of the contact resonance at frequencies given by the bandwidth at half maximum of the contact resonance[36]. The feedback keeps monitoring the amplitude of sidebands and readjusts the frequency of AC voltage. We used Spark 70 Pt tip. The resonance frequency of the tips was ~70 kHz and the force constant was 2 N/m. The contact resonance frequency of PFM was ~250 kHz.

*Optical Characterization of hBN*: The moiré superlattice twisted hBN flakes are mounted on OXFORD MicrostatHires cryostat at 300 K, ~$10^{-6}$ Torr. The samples are irradiated with a 200 kHz pulsed deep-UV laser tuned to 193 nm (~6.42 eV), which is just above the hBN optical band gap 5.76 eV (~215 nm). This output is generated from a laser system consisting of a Yb-doped fiber mode-locked femtosecond laser (Light Conversion Pharos) with 50 $\mu J$ pulse energy at 200 kHz repetition rate and 100-fs pulse width, and UV-capable high power optical parametric amplifier (Light Conversion Orpheus-HP). The beam is injected to the sample at normal incidence with a deep-UV dichroic mirror and focused with a 40x UV-enhanced reflective objective (Thorlabs LMM40X-UVV, 0.5 NA). The incident average power is 149~742 nW and the exposure time is 10s. The photoluminescence emission beam from the sample is gathered in reflection geometry, 193 nm long-pass filtered, and collected to the spectrometer (Princeton Instruments, HRS-500SS, and PyLoN-100BRX).

**Author Contributions**

K.H. and Y.D.K. designed the research project and supervised the experiment. K.H., T.K., and M.C. fabricated the twisted hBN samples. K.H., M.C., S.H.P., and T.K. performed the KPFM measurements and supervised by S.M.Y. and V.M.M. K.H. performed DFRT-PFM measurements. K.H., S.T.L., and S.H.K. performed optical measurements. K.W. and T.T. supplied the hBN crystals. K.H. and Y.D.K. analysed the data and wrote the paper. All authors contributed to the scientific planning and discussions and commented on the manuscript.

**Acknowledgements**

This research was supported by the National Research Foundation of Korea (NRF) grant funded by the Korea government (MSIT) (2021K1A3A1A32084700, 2021R1A2C2093155, 2021M3H4A1A03054856, 2022R1A4A3030766, 2022M3H4A1A04096396, RS-2023-00254055). This work was supported by a grant from Kyung Hee University in 2019 (KHU-20192441). This research was supported by the education and training program of the Quantum Information Research Support Center, funded through the National research foundation of Korea (NRF) by the Ministry of science and ICT(MSIT) of the Korean government (2021M3H3A1036573). S.H.P. and S.M.Y. were supported by NRF of Korea grant funded by the Korea government (MSIT) (NRF-2023R1A2C1003047 and NRF-2022R1A4A1033562). AFM/KPFM and DUV femtosecond laser spectroscopy measurements were supported by the Korea Basic Science Institute (National Research Facilities and Equipment Center) grant funded by the Ministry of Education (2021R1A6C101A437). K.W. and T.T. acknowledge support from the JSPS KAKENHI (21H05233 and 23H02052) and World Premier International Research Center Initiative (WPI), MEXT, Japan. The DFRT-PFM measurements were performed with the assistance of Park Systems. Work at the City College of New York (V.M.M and M.C) was supported by the AFOSR Global grant (FA2386-21-1-4087) and the National Science Foundation (DMR-2130544).

# Figures

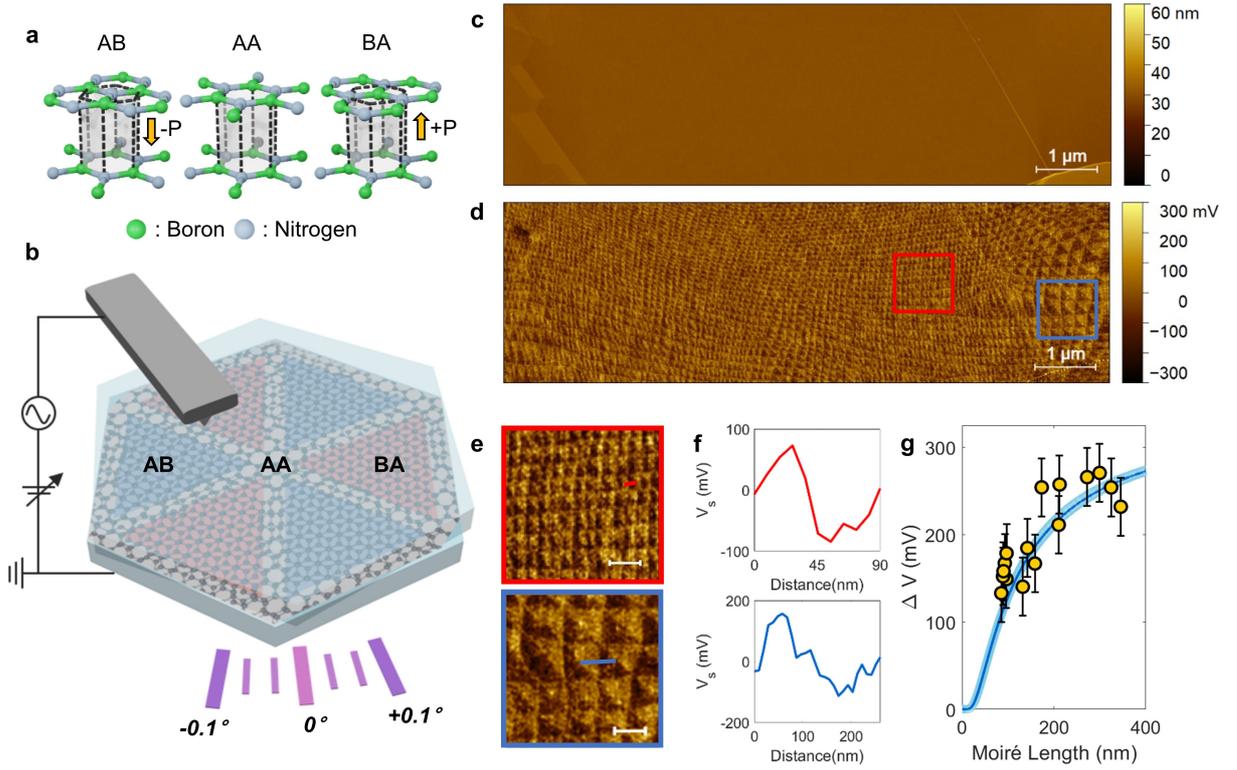

**Figure 1 | Moiré superlattice potentials from twisted hBN. a**, The atomic structures of stacking orders in twisted hBN and the moiré pattern reconstructed configuration. The AB, AA, and BA stacking orders have different electrostatic potential values because the dipole moments can create the polarization along out-of-plane direction. In AB (BA) stacking, the polarization direction is downside (upside). Since AA stacking is unstable, the atomic relaxation leads to triangular shape AB/BA moiré patterns. **b,** Schematic of KPFM measurement of twisted hBN. **c,** AFM topography image of a representative sample. **d,** Corresponding KPFM image demonstrating regular moiré superlattice potentials. **e,** Zoom-in KPFM images in **(d)** (all scale bars = 200 nm) **f,** Potential line profiles extracted from red and blue lines in **(e)**. The red (blue) line is from the smaller (larger) moiré pattern. **g,** Surface potential difference by the size of moiré length. The yellow dots are representative data collected from **(d)** and the blue line is fitted with $\Delta V \sim exp(-\frac{4\pi z}{\sqrt{3}\lambda_m})$ from equation (1).

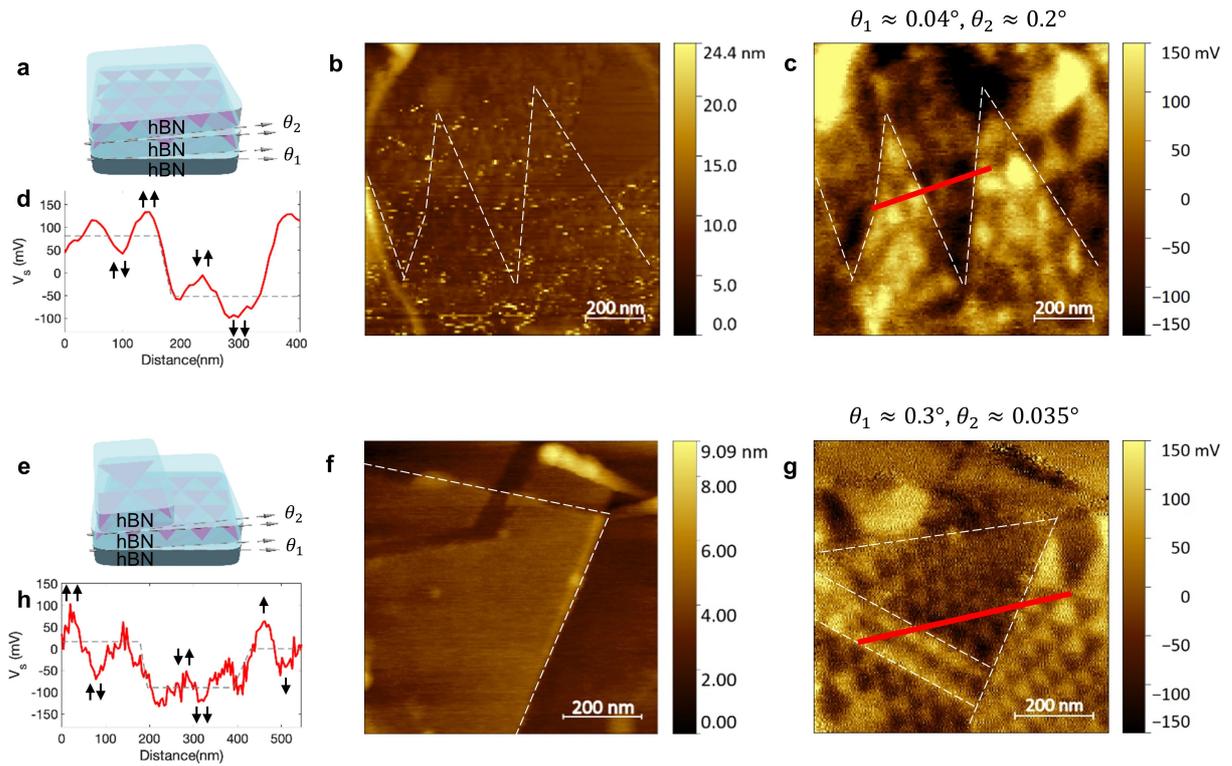

**Figure 2 | Cumulative hBN moiré superlattice potential by additional interfaces. a**, Schematic of twisted hBN with two different superlattice interfaces. **b**, AFM topography image of the sample. **c**, Corresponding KPFM image. It has twisted angles ($\theta_1 \simeq 0.04°$, $\theta_2 \simeq 0.2°$) from two different interfaces. Whited dashed lines show the bigger moiré pattern boundary. **d**, Potential line profile in **(c)** demonstrating two level average potentials due to the superimposition of two different moiré systems. **e**, Schematic of twisted hBN which has both single interface and two interfaces simultaneously. **f,** AFM topography of the sample. White dashed lines show the additional top hBN layer on the twisted hBN. **g**, Corresponding KPFM image. It has twisted angles ($\theta_1 \simeq 0.3°$, $\theta_2 \simeq 0.035°$) from two different interfaces. The moiré domain by additional layer is encircled by white dashed lines. **h**, Potential line profile in **g** originating from one interface system and two interfaces system. It also shows the step-like increase and decrease because of the multi-ferroelectric polarization.

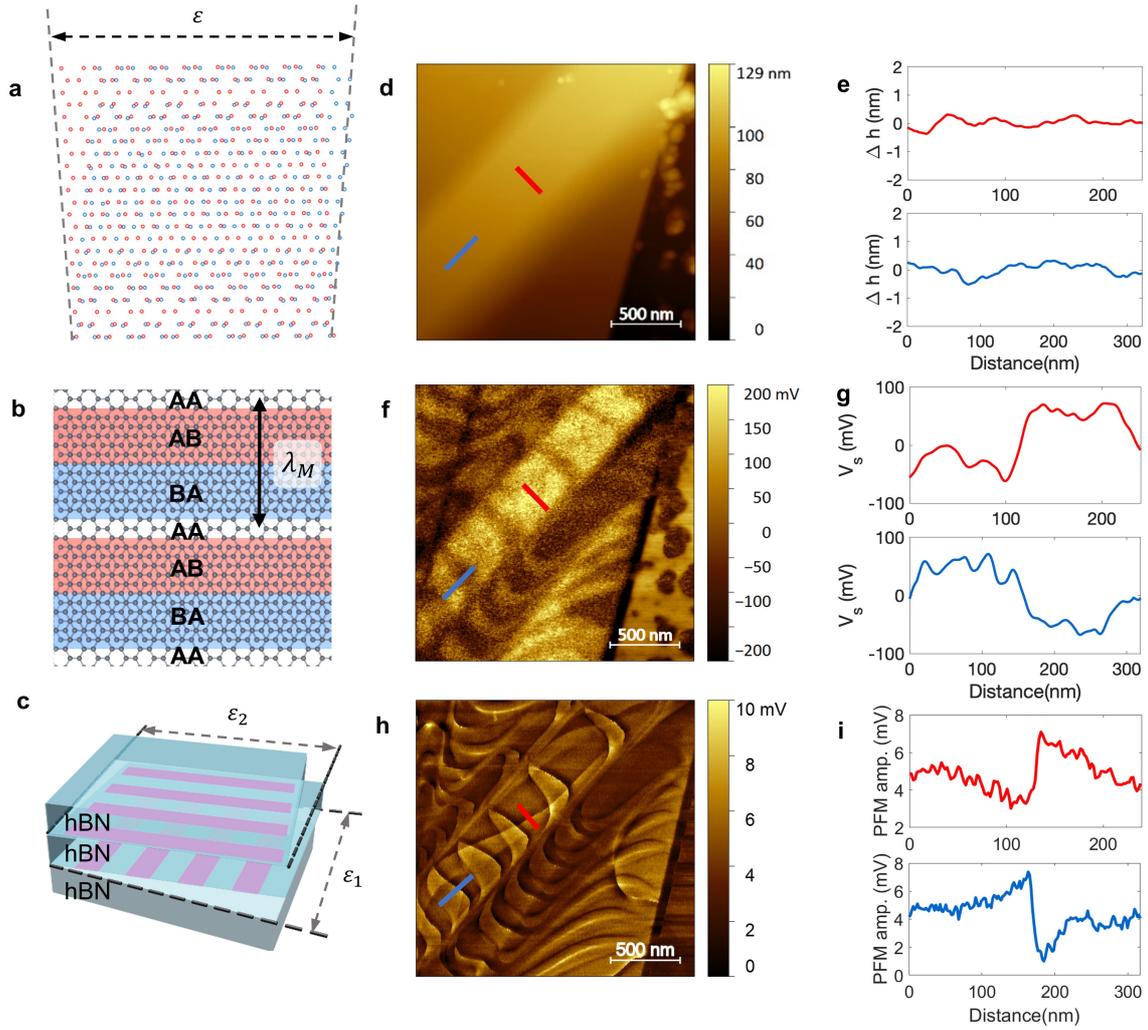

**Figure 3 | Anisotropic moiré domain induced by horizontal simple shear. a,** Atomic structure of the horizontal simple shear (HSS) when two layers are mismatched by strain. **b**, Atomic configuration after relaxation. **c**, Schematic of HSS hBN with two different superlattices. **d**, AFM topography image of a representative sample. **e**, The line profiles from (**d**) showing flat surface. **f**, Corresponding KPFM image. **g**, The line profiles of surface potential in (**f**) from the first superlattice layer (red) and the second layer (blue). The 1-dimensional moiré length ($\lambda_m = a/\varepsilon$) is inversely proportional to $\varepsilon$. Since moiré lengths from the first (second) superlattice is $\lambda_{M1} \simeq 240$ nm ($\lambda_{M1} \simeq 320$ nm), it is simply sheared by $\varepsilon_1 \simeq 0.104\%$ ($\varepsilon_2 \simeq 0.078\%$). **h**, DFRT-PFM image of the same area and (**i**) its line profiles.

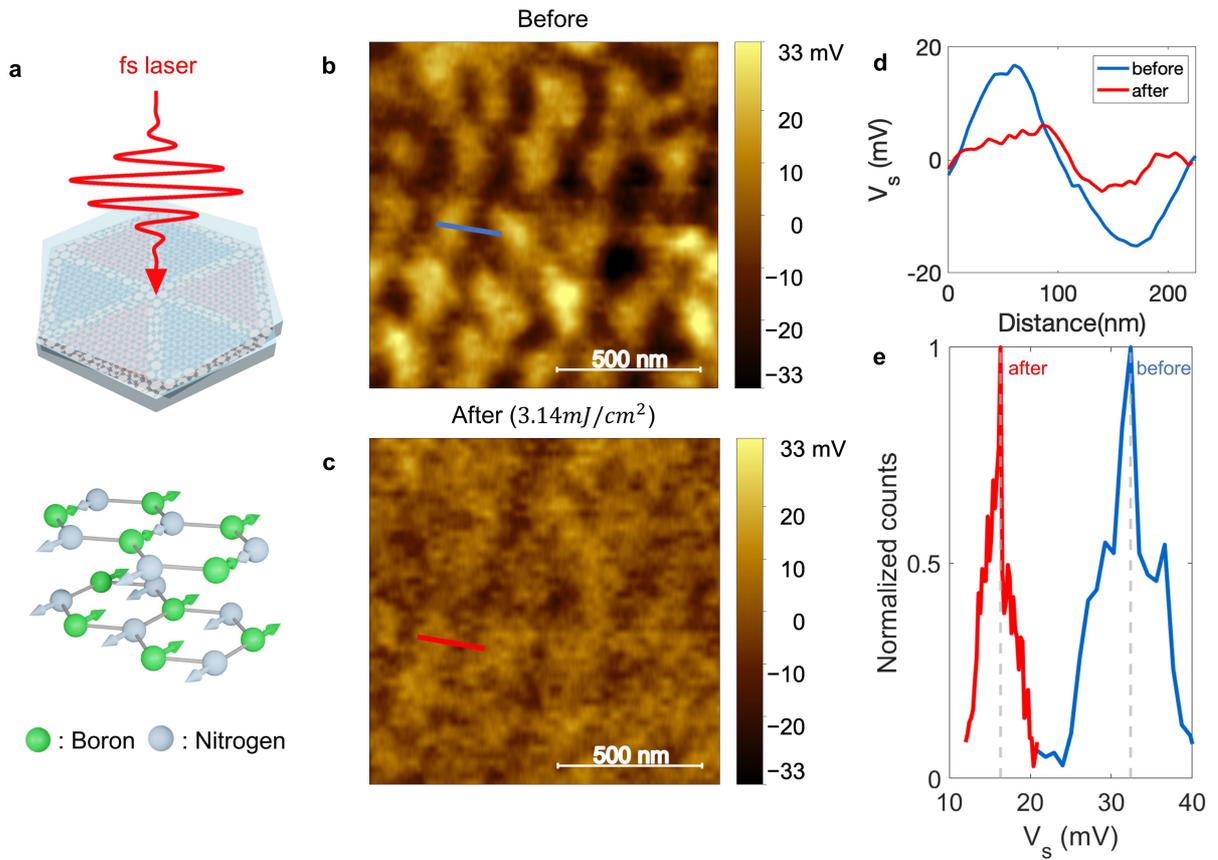

**Figure 4 | Femtosecond laser-induced *in-situ* engineering of moiré superlattice potential strength**. **a,** Schematic of femtosecond pulse laser irradiation to twisted hBN moiré platform. Due to the strong optical phonon coupling, boron and nitrogen can be deformed in opposite directions. The motions of the stacked BN planes relative to each other give rise to rigid-plane shear and breathing/compression modes at low energy. **b-c,** KPFM image of a representative sample **(b)** before and **(c)** after pulse laser irradiance of 3.14 mJ/cm$^2$. **d,** The line profiles of moiré superlattice potential along the blue and red line in **(b)** and **(c)**. **e,** The normalized number of counts of moiré potential depth data collected in **(b)** and **(c)**.